\documentclass{jaa}
\usepackage{graphicx}
\usepackage{url}
\usepackage{xcolor}
\usepackage{array}
\usepackage[colorlinks=true,citecolor=blue]{hyperref} 
\usepackage{booktabs}
\usepackage{soul}
\usepackage{enumitem}
\newcolumntype{C}[1]{>{\centering\arraybackslash}p{#1}}

\newcommand{\aap}{    {\it A \& A}}

\newcommand{\aj}{     {\it AJ}}
\newcommand{\apj}{    {\it ApJ}}
\newcommand{\apjl}{   {\it ApJL}}

\newcommand{\mnras}{  {\it MNRAS}}

\newcommand{\arcmin}{^{\prime}}
\newcommand{\arcsec}{^{\prime\prime}}
\newcommand{\et}{\textit{et al.}}

\begin{document} \sloppy

\title{An overview of the proposed INdian Spectroscopic and Imaging Space Telescope (INSIST)}

\author{Annapurni Subramaniam\textsuperscript{1,*}}
\affilOne{\textsuperscript{1}Indian Institute of Astrophysics, Sarjapur Road, Koramangala, Bangalore, India.\\}

\twocolumn[{

\maketitle

\corres{purni@iiap.res.in}

\msinfo{18 December 2021}{TBD}

\begin{abstract}
India reached a major milestone in the area of space astronomy with the successful launch and post-launch operations of its first space observatory, AstroSat.  The success of this space observatory and the lessons learned must be utilized effectively to enlarge the footprint of Indian space astronomy in the international scene. In response to a call for proposals by the Indian Space Research Organisation, a detailed proposal for a next generation UV-optical mission, the INdian Spectroscopic and Imaging Space Telescope (INSIST) was submitted. Combining a large focal area with a simple and efficient optical design, INSIST is expected to produce HST-quality imaging and moderate resolution spectra of astronomical sources. The main science drivers for this mission span a wide range of topics, starting from evolution of galaxies in groups and clusters, chemo-dynamics and demographics of the nearby universe, stellar systems with accretions, to stars with planetary systems, to cosmology near and far. The proposal was awarded seed funding and has completed two years of pre-project phase. An overview of this proposed mission is presented here along with the current status.
\end{abstract}


}]


\doinum{12.3456/s78910-011-012-3}
\artcitid{\#\#\#\#}
\volnum{000}
\year{0000}
\pgrange{1--}
\setcounter{page}{1}
\lp{1}

\section{Introduction}

Optical imaging has been the driving force of astrophysical research, historically started by the ground-based telescopes.  Ultraviolet (UV), X-ray and higher energies detected primarily from space, and infra-red (IR) from ground and space added to the multitude of discoveries in recent times. 
Multi-messenger astronomy have proved that multi-wavelength is indeed the powerful way to explore the Universe. In the next decade, the astronomy will enter the era of large scale surveys in the optical, with the 
Legacy Survey of Space and Time (LSST) by 8-m Vera C. Rubin Observatory,
apart from other survey programs, mostly in the optical and near-IR pass bands. The next generation ground observatories have to be supported by the next generation space observatories to cover the multi-wavelength domain to get into the realm of unknown-unknown.

In the international scene, apart from the James-Webb Space Telescope (JWST) launched in 2021, the other planned missions in the optical to IR wavelengths are EUCLID (Euclid is a visible to near-infrared space telescope currently under development by the European Space Agency (ESA) and Nancy Grace Roman Telescope (formerly WFIRST) by NASA. The main science goals of these missions are to understand dark energy, along with a few well-defined legacy value science cases.  In the ground, the LSST is expected to begin its operations very soon, followed by a decade long schedule of observations, with the data release of the deep observations expected to be starting from around 2030. The high resolution optical/UV images have so far been provided by the Hubble Space Telescope (HST) since 1990s.  The high spatial resolution imaging of the HST is limited to very small fields. As this mission is approaching the end of its life-time, it is unlikely to be available for a very long period. Galaxy Evolution Explorer (GALEX) provided the UV sky survey from 2003--2013. As the 2030s will produce deep sky images in the optical and IR regions, a similar coverage in the UV are not addressed by any planned mission. This is where India, with its capacity to launch and operate space observatories, can fill the gap.

India is also now ready for a next-generation mission, following the success of the AstroSat, particularly in the UV region. The Ultra-violet Imaging Telescope (UVIT) was indigenously made starting from design to fabrication, integration, test and calibration with the IIA as the lead institute in collaboration with several Indian Institutions such as Inter-University Centre for Astronomy and Astrophysics (IUCAA), Tata Institute of Fundamental Research (TIFR), several institutions under Indian Space Research Organisation (ISRO), and the Canadian Space Agency (CSA) (Kumar \et\ 2012). The UVIT was a capability demonstration, and it is important to take this strength to create the next generation UV-optical observatory class space telescope for the next decade, to fill the gap in the international scene. \\

In the last 5 years, a number of important science results have come out using the UVIT data, that are led by the Indian astronomers. Some of the important one include, the discovery of a Lyman continuum galaxy at a red-shift of z=1.42 (Saha \et\ 2020), the discovery of a large, diffuse star forming galaxy (Yadav \et\ 2022), discovery of binary mass-transfer pathway for the formation of blue straggler stars in star clusters (Sindhu \et\ 2019, Singh \et\ 2020, Jadhav \et\ 2021, Pandey \et\ 2021, Vaidya \et\ 2022), star formation in the tidal tails of the merged galaxy, NGC 7252 (George \et\ 2018a, 2018b), UV bright  horizontal branch stars in globular clusters (Subramaniam \et\ 2017, Rani \et\ 2021, Kumar \et\ 2021), etc.). 

\begin{figure}
  \centering
 \includegraphics[width=0.35\textwidth]{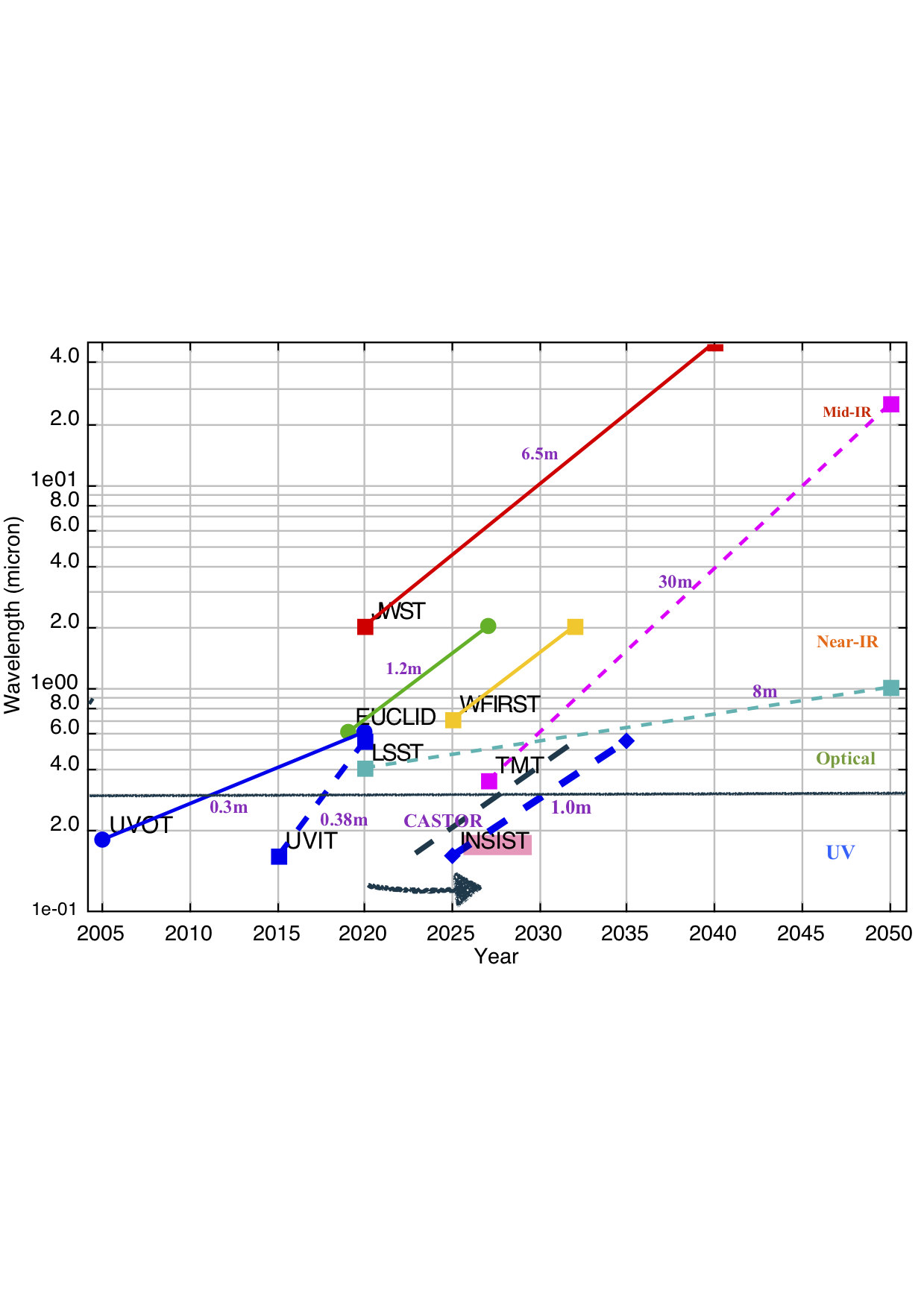}
    \caption{The international landscape of planned ground and space based facilities/observatories covering UV to IR wavelength. This figure was part of the submitted proposal.}
  \label{fig:landscape}
\end{figure}

It was also clear that there is going to be a lack of the UV missions of this type in the next decade, as shown in Figure \ref{fig:landscape}. The ground and space missions that are approved are shown in this diagram along with the year of its commissioning (as per the data available at the time of the proposal submission time). India can utilise this opportunity and can play an important role to fill this gap in the next decade. 

This article presents an overview of the next generation space telescope, the INdian Spectroscopic and Imaging Space Telescope (INSIST).

\section{Planning for a future mission}

The initial days after the launch of the AstroSat and opening of the doors of the UVIT were dedicated to the in-orbit calibration of the Far-UV, near-UV channels, till March 2016, followed by the guaranteed time observations. The UVIT instrument team was initially focused on streamlining the data pipeline to create science ready images, calibration outputs including various tables and images that will be used in the data pipeline, as well as understanding the overall performance of the UVIT (Tandon \et\ 2017, 2020). 

The first science meeting of the UVIT was organised on 6--7 July 2017 in IIA. The meeting was meant to present the calibration, data pipeline and data products from the Instrument team, followed by early science results. As it was getting clear that the UVIT was performing better than expected, particularly in terms of the spatial resolution of $\sim1.5\arcsec$ (better than the design value of 1.8$\arcsec$), it was important to take a look at the future prospects of this achievement. The last session of the meeting was dedicated to the future directions. In this session, there was a presentation by the IIA team on an imaging and multi-object spectroscopic optical-UV mission, capable to address a number of science questions, for which the UVIT has given a headway. The highlights were the high-resolution imaging capability of about 0.2$\arcsec$ in the UV/optical wavelengths and a moderate resolution spectrograph in the UV. This idea was well received by the community present in the meeting. The meeting held to celebrate the second anniversary of the ASTROSAT launch also had a session on future missions. It was suggested to put out an announcement of proposal for future astronomy space missions by ISRO.

In January 2018, the Announcement of Opportunity (AO) for submitting proposals for future astronomy missions was made public by ISRO. The IIA led the efforts to put together a proposal in partnership with other institutions in the country, such as IUCAA, TIFR, CHRIST (deemed to be university), PRL and Aryabhatta Research Institute of Observational Sciences (ARIES). A possible international collaboration with the CSA and the Canadian mission, CASTOR (Cosmological Advanced Survey Telescope for Optical and ultraviolet Research) was also envisaged. The proposal for the future mission, named as INdian Spectroscopic and Imaging Space Telescope (INSIST; PI: Annapurni Subramaniam) was submitted to ISRO in April 2018. A committee set up by the ISRO reviewed all the submitted proposals and found 4 proposals, including INSIST, to have the maturity to go through a pro-project phase. The advisory committee for space (ADCOS) recommended INSIST proposal for a pre-project phase with a seed funding. Thus, the INSIST proposal progressed to a seed funded pre-project phase in March 2019.

\section {Details of the INSIST mission}

The top level science-driven requirement for INSIST, is a large focal area with an efficient optical design, to produce HST-quality imaging and moderate resolution slit-less as well as multi-object spectra of astronomical sources to address science questions that span a wide range of topics. 

It is therefore proposed that, INSIST, an observatory class mission, will have a high resolution imaging  ($\sim0.2\arcsec$) capability using a 1-m primary mirror, to achieve a detection limit of $\sim$26 magnitude in the UV and $\sim$29 magnitude in the optical, for an exposure time of 1 ksec. A moderate resolution multi-object spectrograph (MOS) (R$\sim$1000--2000), capable of obtaining spectra up to a limiting magnitude of 20 in UV, is also proposed for objects falling within a field of $\sim3\arcmin\times3\arcmin$. The proposed design of the INSIST has a relatively low risk and is technically feasible. The payload is designed for the extended PSLV fairing, with an appropriate mass and volume. We envisage this telescope to be completed and launched in 7 years, once it is approved. INSIST is proposed as a collaborative mission with several Indian Institutions. A strategic collaboration with CSA is envisaged for some specific sub-systems of the project.

\par

\subsection{Main Science Goals}

The science goals for this mission have a large depth and breadth, with the ability to cater to a wider science community in the country that is already trained during the UVIT/ASTROSAT phase. The science document put together for the proposal is thus a natural outcome of the wish list of science cases brought out by the user community. The  instrument specifications were derived from the science cases and were used to finalise the specifications of INSIST. The proposed science goals will be carried out by this mission as legacy surveys of a few down selected topics, in combination with a Guest Investigator program. 
This approach is expected to address specific science goals, that would leave behind a legacy value for this mission in the international scene, as well as to efficiently use the observing time of the mission -- a valuable lesson learned from the ASTROSAT operations. 

\begin{figure}
  \centering
 \includegraphics[width=0.48\textwidth]{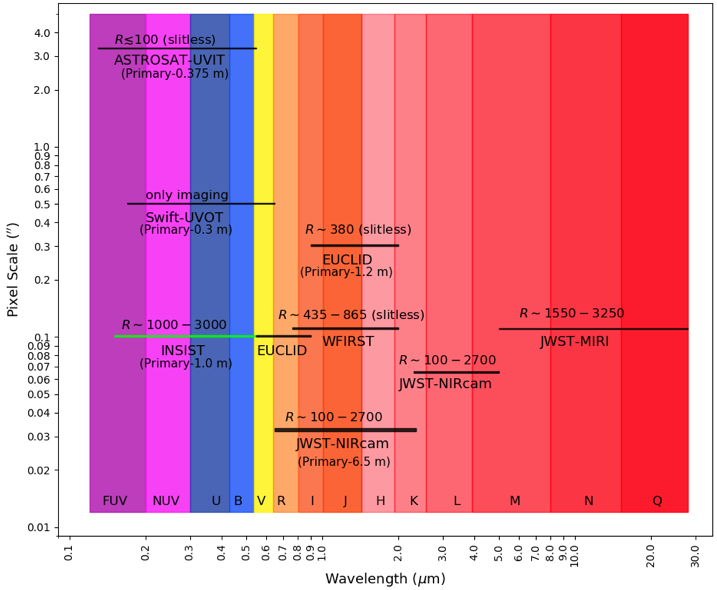}
    \caption{ The spatial resolution is shown against wavelength for INSIST and other missions expected to be operational in the next decade. This figure was part of the submitted proposal (Figure Courtesy -- Maheswar Gopinathan).}
  \label{fig:spa_res}
\end{figure}

The expected spatial resolution of INSIST as a function of wavelength is shown in Figure \ref{fig:spa_res}. The plot compares the location of INSIST in this plane, with respect to the other missions which will be operational in the next decade. The complementary nature of this mission in the UV, with similar resolution, brings out not only the synergy with other missions, but also the ability to address some outstanding science cases. The dramatic improvement in spatial resolution, in combination with the deep detection limit in the UV, is expected to make this mission capable of addressing a wide range of science cases, across all topics in astrophysics.

The main science cases for the proposed mission addressed using imaging and spectroscopy are:
\begin{enumerate}[noitemsep]
\item Galaxy formation and evolution 
\item Galaxy clusters: Modes of triggering and quenching of star formation
\item High red-shift galaxies – active galactic nuclei (AGNs), Quasars
\item Resolved young stellar population in the Local Group 
\item Chemo-dynamics and demographics of resolved stellar population 
\item Globular cluster populations
\item Mapping the magnetospheres of young stellar objects (YSOs), feedback from proto-stars
\item Chromospheric activity and planetary systems
\item Time-domain astronomy and the variable universe
\item Cosmology near and far
\end{enumerate}

We grouped the above science goals along with their requirement for instrument specification, in Table \ref{tab:sci_req}).

\begin{table*}
    \centering
    \begin{tabular}{llllll}
    \toprule
    No. & Science& Sub-topics & FoV & Imaging       & Spectral \\
        & Theme  &            &     & resolution    & resolution\\
    \hline
    1 & Stellar & Magnetospheres & 5-10$\arcmin$ &0.2-0.4" & $\sim$2,000+, multiplexing \\
 & & Planet formation in YSOs & &&\\
 & &Mass Assembly of nearby galaxies  & &&\\
 & & Resolving the cores of GCs  & &&\\
 & &  IMBH, LMXBs and CVs & &&\\
 & & planet hosting stars& & & \\
 \hline
 2 & Galaxy & Census of Dwarf galaxies &10-15' &0.2-0.4" & $\sim$1000+, large multiplexing \\
 & &Interacting systems  & &&\\
 & & All types of galaxies & &&\\
 & & TDGs, tidal streams & &&\\
 \hline
 3 & Dense systems & Nuclei of galaxies &$<$ 10' &0.2" & $\sim$1000+, spatially resolved  \\
 & & AGNs and Quasars & & & multi-object spectra\\
 & & SFRs in galaxies &&  & \\
 & & Dense star clusters &&& \\
 \hline
 4 & Galaxy Clusters & SF in deep fields &10-15' &0.2" & $\sim$1000+, large multiplexing \\
 & & Jelly fish galaxies &&  & \\
 & & tidal features and SF &&  & \\
 & & merging galaxies &&  & \\
 && SF galaxies in cosmic noon & & & \\
 \hline
 5 & Time Domain & Supernovae, TDRs & 5' & 0.2" & R=500-1000 \\
 & & GW counterparts &&  & with good sensitivity\\
 & & Various types of Outbursts & & & \\
    \bottomrule
    \end{tabular}
    \caption{Science goals of INSIST and the required instrument capabilities. }
    \label{tab:sci_req}
\end{table*}

The science goals and specifications from the science team as captured in Table \ref{tab:sci_req} is then used to arrive at the specification of the INSIST telescope and the back-end instruments.

\subsection{INSIST specifications}
The requirements of a mission are generally science driven and the preliminary requirements for the INSIST were derived based on the science requirements, as shown in Table \ref{tab:sci_req}. The impact of this mission in UV, optical imaging and spectroscopy will depend on its matching performance and complimentary with other  surveys that are expected to be available in the next decade. There is a demand for high-resolution imaging, similar to the HST, for a larger field of view so as to produce the spatially resolved images of dense star clusters, nuclei of nearby galaxies, star forming regions etc. It is also required to bring the sensitivity and resolution at par with the various surveys that will be available from ground and space observations in the next decade, in the other wavelength region. Similarly, the depth of the UV/optical imaging should be adequate to be able to combine the data with other surveys.

The cost of a mission will primarily depend on the size of the primary mirror and the focal plane array. In this mission, the size of the primary mirror was mainly decided based on the expected depth of observations and resolution. Other factors that dictate the size would be the requirement of laboratory space for integration of the entire payload, availability of suitable optics and coating facilities as well as a launch vehicle for the size of the payload.

\begin{figure}
  \centering
 \includegraphics[width=0.48\textwidth]{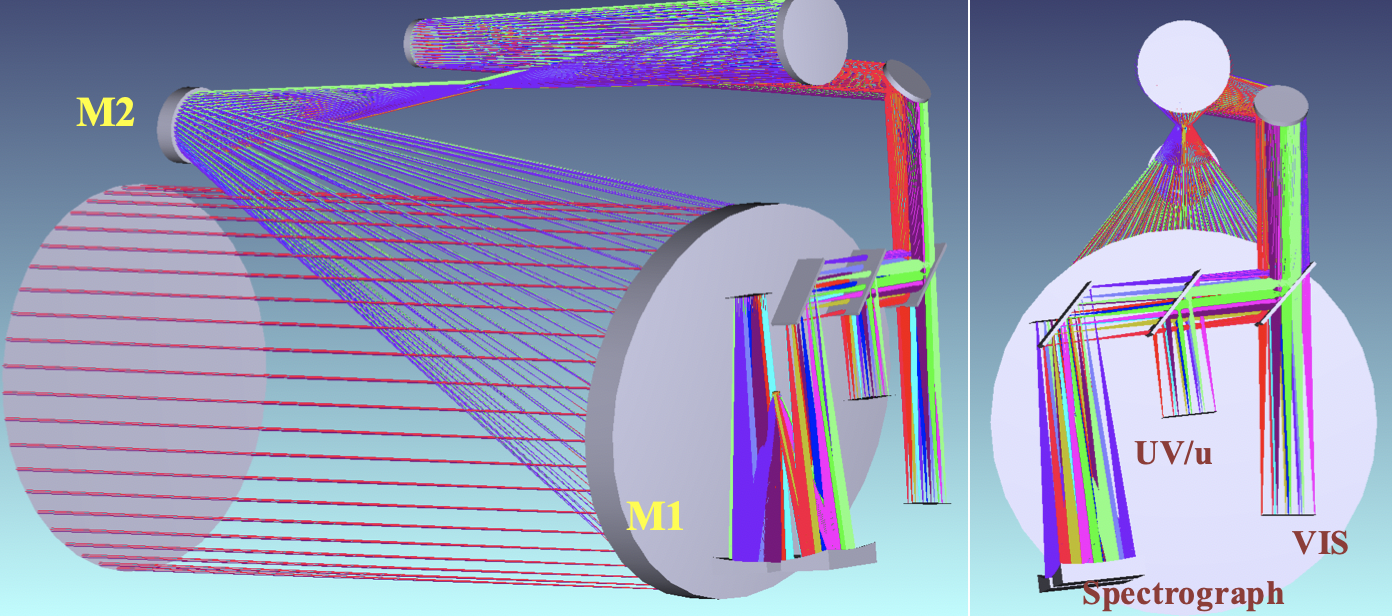}
    \caption { One of the optical designs proposed for INSIST. This is taken from the submitted proposal. There are three imaging channels, but two are shown in the figure. The UV/u option is further split to separate the two, using a beam splitter. This is a TMA design with 1-m off-axis primary mirror. }
  \label{fig:opt_design}
\end{figure}

\begin{figure}
  \centering
 \includegraphics[width=0.48\textwidth]{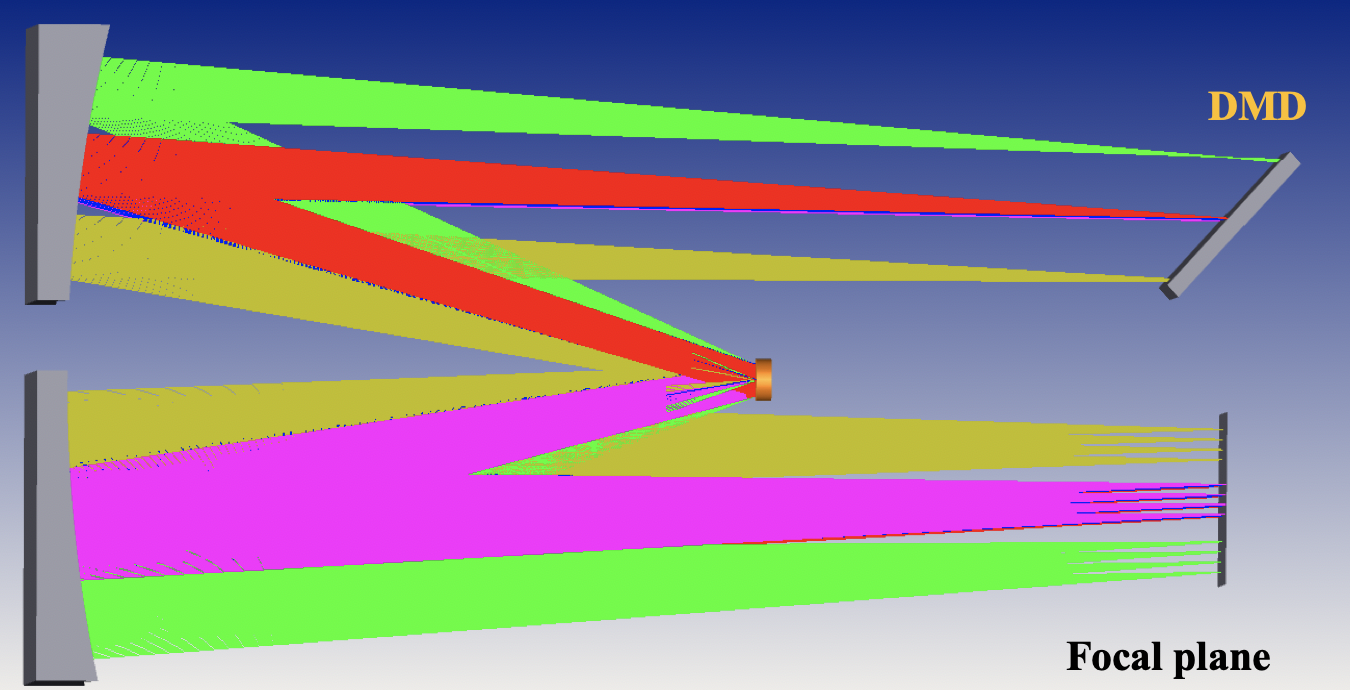}
    \caption { One of the optical designs proposed for the DMD-MOS on INSIST. A convex grating is used in this design to disperse the beam. This is taken from the submitted proposal.}
  \label{fig:mos_design}
\end{figure}

The preliminary specifications as submitted in the proposal is listed in Table \ref{tab:INSIST-spec}.  INSIST is proposed to be a 1-m aperture telescope with an f-number of 20. It is proposed to have simultaneous imaging in the UV, u and g bands with a large field of view. The UV and u channels are proposed to have slit-less spectroscopy as well. A parallel and a smaller field will be used for multi-object spectroscopy (MOS) in the UV. The MOS will be achieved using a digital micro-mirror divide (DMD) that is capable of sampling the sky at 10$\mu$m level (0.2$\arcsec$).  The proposed initial concept design for the optical design is shown in figure \ref{fig:opt_design} and that for the DMD based multi-object spectrograph is shown in \ref{fig:mos_design}. Some of the parameters have been modified since then, and these will be presented in Sriram \et\ (2022, under revision).

\begin{table*}
    \centering
    \begin{tabular}{lllll}
    \toprule
Description & UV & u & g & UV-MOS\\
\hline
Wavelength (nm) & 150--300 & 300--400 & 400--500 & 150--300 \\
Field of view (degree) & 0.5 & 0.5 & 0.5 & 0.1 \\
Spatial resolution & 0.2" & 0.2" & 0.2" & 0.2" \\
Sensitivity (1 ksec)& $\sim$26mag  & $\sim$26 & $\sim$29 & $\sim$20 mag \\
Multiplexing & - & - & - & 200 \\
Filters & TBD & TBD& TBD& -\\
Spectral resolution & $\sim$100& $\sim$100&-& 1000 - 2000\\
Orbit&\multicolumn{4}{c}{Sun-synchronous} \\
    \bottomrule
    \end{tabular}
    \caption{The preliminary specifications of INSIST. The UV, u and g bands will have imaging and provision to obtain slit-less spectra. These fields will be imaged simultaneously by three focal-plane detector arrays. The UV-MOS is a UV multi-object spectrograph with only spectroscopic capability, of an adjacent field with a FoV of a few arcmin. The filters that are required for the various channels are to be decided.  }
    \label{tab:INSIST-spec}
\end{table*}

The AstroSat has an equatorial orbit with 90 minute to complete one orbit. In each orbit, the night sky is available only for  one-third of the orbit in favourable cases. This limits the efficiency of observation per orbit. The proposed orbit for INSIST is a Sun-synchronous orbit to provide an efficient use of the time in orbit.

\section{Participating Institutions}
This mission is proposed as a mission led by the Indian Institute of Astrophysics, in collaboration with several Institutions in the country. The following Institutions are mentioned as partners in the submitted proposal. The technical as well as the science teams were drawn from these institutions as well.

\begin{itemize}[itemsep=0.01cm]
    \item Lead Institute: Indian Institute of Astrophysics
    \item Collaborating Institutions: TIFR, IUCAA, Physical Research Laboratory (PRL), ARIES and CHRIST (deemed to be University)
    \item ISRO: Various centres of ISRO will collaborate in development of certain sub-systems and science. The Launch and operations of the mission will also to be supported by ISRO. 
    \item International partner: The Canadian space Agency and the CASTOR mission team are expected to be partners for this mission.
\end{itemize}

Most of the major astronomy institutions with UV-optical expertise are part of the proposal. It is envisaged that new collaborating institutions will join the project as it moves forward.

\section{A summary of the activities so far}

The pre-project phase of the proposal is expected to look into these critical sub-systems along with the overall optical and mechanical design maturity of this mission. A few trade studies to understand various sub-systems are also to be performed as part of the pre-project phase.  The pre-project phase of INSIST mission started in March 2019, with the release of the seed funding.  The optical design has matured significantly (Sriram et al. 2022, under revision) and the mechanical design also has progressed considerably. There are a few critical subsystems in the proposed INSIST mission, that require detailed technical evaluation.

The activities that are being carried out as part of the pre-project phase include the following:
\begin{itemize}[itemsep=0.01cm]

    \item {\bf DMD:} The procurement and experimentation with the DMD to (1) understand the functionality, (2) develop capability to command the micro-mirror movements and (3) characterise its performance for the MOS. The DMD array, that is ideal for the above purpose was identified and was procured along with the control electronics. Currently, the team is able to address and send commands to individual micro-mirrors. The team is working on developing a simpler control electronics for the DMD to be used in the MOS.
    
    \item {\bf DMD based MOS:} The MOS with a spatial resolution of 0.2$\arcsec$ with a sampling size of 10$\mu$m cannot use slits. This sampling can be achieved by using a DMD. So far the DMD is being tested in the laboratory for integrating into a spectrograph. The UV properties of the micro-mirrors of the DMD also need to be assessed.  Demonstration to use DMD in a MOS in a laboratory set up is being carried out. 
    The micro-mirrors get tilted to a specific angle when they are commanded and this helps in reflecting light from specific area to go through the spectrograph. Currently, a lab model of the spectrograph with the required optics is almost complete, along with the DMD and cameras to acquire images as well as spectra. Detailed tests are to be performed soon to understand the performance and the characteristics of spectra and images obtained by this set up.

   \item Spatial resolution: The most important aspect of the proposed mission is the high spatial resolution of 0.2". This is expected to be achieved by an active real-time image correction performed by a fast-steering mirror (FSM). The plan is to use a region-of-interest (RoI) in g-band (VIS) images to estimate the image motion and correct it using an FSM. This technology is available and was one of the contribution to the JWST mission from the CSA. The FSM will be able to handle the image motion correction, within a specified limit of the spacecraft drift and jitter. For the INSIST mission, the FSM is considered to be one of the deliveries from the Canadian partner. Detailed technical feasibility study is currently being carried out by CSA with their Industry partners in Canada.
   
    \item {\bf Partnerships:} There was a one-day meeting of the INSIST national team in IIA in September 2019. 
    The national partnership details were discussed and a preliminary list of roles of the collaborating institutions were put together. This was followed by a
     3 day meeting of the INSIST and CASTOR teams. In the meeting, it became clear that both the teams are willing to work together and look forward to a joint mission. The teams decided to work together to come up with a common optical design first, then proceed to further steps. Regular meetings are help between the national teams (bi-weekly). The INSIST and CASTOR technical teams have been working together for a joint optical design.
     
     \item Focal plane array: This mission proposes to have simultaneous imaging in three bands, that requires three focal plane arrays. The space qualified detectors for the UV requires coating and the electronics is a major component of the mission. There is a requirement to keep the dark current and read noise to be as minimum as possible to detect faint sources. Trade studies to identify and test the performance of available detectors is being carried out by CSA.
     
     \item Data volume: The large field of view and three simultaneous focal plane arrays are expected to produce a large amount of data for download. The meta-data, the engineering data and the science data put together will have be a  large volume of data. The high-speed/efficiency data download is another requirement for this mission. An optimal data compression also need to be worked out for efficient data download.
     
     \item {\bf Engaging with the community:} An invited talk on INSIST was presented in the special session titled, "Future Astronomy Payloads / Missions",  in the 38th meeting of the Astronomical Society of India, held in IISER, Tirupati. A one-day workshop titled "Ultraviolet space astronomy: UVIT and beyond" was conducted by the INSIST team, as part of the 39th Annual meeting of the Astronomical Society of India, on 18th February 2021. The aim of the workshop was to let the scientific community of the country know about the proposed INSIST mission, and its details. The workshop was attended by 97 participants from various parts of the country for 28 different institutions. A total of 21 speakers made presentations during the workshop. A panel discussion was organised towards the end of the workshop where INSIST team initiated a discussion on various aspects of the mission ranging from science to telescope and instrument specifications, and urged  the participants to reach-out to the INSIST team in case if they are willing to participate and contribute especially in science data simulations and in the development of various software tools for testing and calibration. The you tube link to the proceedings of the meeting is available in the website of ASI-2021 meeting (https://astron-soc.in/asi2021/home). 
    
    \item {\bf Science document:} The national team worked on a detailed science document, expanding on the initial science document that was submitted to ISRO, along with the proposal. The expanded science document was submitted to ISRO a year ago, and we received excellent feedback. The science team members are currently working on putting together a few manuscripts for publication, based on the science document. Therefore details of the science cases are not presented here.
    
    \item {\bf Reviews:} During the pre-project phase, that has now completed about two-and-a-half years, the progress was reviewed four times by a standing committee set up by ISRO. The review committee provided excellent and positive feedback to help complete the objectives and to move forward in the project. The team looks forward to moving to the next phase of this project.
\end{itemize}

\section {Discussion}

 In the national scene, the Indian astronomical facilities in the present decade witnessed a boost with the commissioning of the 3.6-m Devasthal Optical Telescope (DOT) (Sagar \et\ 2019), the upgrade of the Giant Metrewave Radio Telescope (GMRT; (Gupta \et\ 2107), access to the SALT 8-m telescope along with the successful launch and operations of AstroSat. The next decade will see the full operation of the DOT and the commissioning of the 2-m class telescope of the Physical Research Laboratory (PRL). India also plans to be a partner in the LSST operations, which will give access to the optical synoptic survey data as well as deep photometric data. The Thirty Meter Telescope with the 10\% Indian partnership is expected to see the first light by 2030s. These two will give access to the world-class facilities for the Indian astronomical community in the next decade, in the optical and IR wavelengths. A complementary Indian facility in space for UV-optical imaging and spectroscopy will pole-vault the Indian capability and its footprint in the next decade.

 AstroSat project which took more than a decade to realise, generated a lot of know-how regarding this niche area of engineering and technology in the country (Seetha \& Kasturirangan, 2021). A large pool of trained resource is available in the country today in various aspects of making and operating a space telescope, which include Optics, Mechanical, electronics, data processing, operations as well as various aspects related to mission control and commanding. The MGK Menon Space Sciences Laboratory at the Centre for Research and Education Science and Technology (CREST) campus of IIA was set up for the Assembly, Testing and Integration of the UVIT payload. This state-of-the-art facility is presently used for the integration of the Visible Emission Line Coronagraph (VELC) to be flown on the ADITYA, the Solar L1 mission (Raghavendra Prasad \et\ 2017).
 
 The members of the INSIST team are drawn from the AstroSat/UVIT mission. The infrastructure in the MGK Menon Laboratory will be available once the VELC is delivered and operational. It is therefore necessary that the infrastructure, and technical and scientific human resources created in the country in the last two decades it taken to the next level. We hope that the INSIST mission goes forward to the Phasa-A stage shortly and becomes a project in the next couple of years.

\textbf {Summary:}

This article presents an overview of the proposed next generation UV-optical space telescope, INSIST. India has the advantage of the successful launch and operations of the AstroSat, that has completed 6 years in orbit. The experience, infra-structure, human resources as well as the know-how that was developed for UVIT over the last two decades should be put to use by planning and executing a next-generation UV-optical space telescope.  

The INSIST mission proposal was submitted to ISRO in April 2018, with IIA as the lead institute and other collaborating institutions within the country. The pre-project phase started in March 2019 and a seed funding was provided to demonstrate a few critical components. The optical design, partnerships, detailed science proposal, procurement and control of the DMD and a lab model of the DMD based MOS, engagement with the community are some of the activities that have been completed during the pre-project phase. The team looks forward to the Phase-A study phase as the next step.




\section*{Acknowledgements}
This article is on behalf of the entire INSIST team and the author thanks and acknowledged all the members for the team for their contribution. Seed funding for the pre-project phase is gratefully acknowledged. Thanks to the referee for the comments that helped to improve the manuscript. 

\vspace{-1em}


\bibliography{references} 

\end{document}